\documentclass[twocolumn,amsmath,floats,showpacs,nofootinbib]{revtex4}
\usepackage[usenames]{color}
\usepackage{graphicx}% Include figure files
\usepackage{epstopdf}
\usepackage{textcomp}
\usepackage{hyperref}
\usepackage{multirow}
\usepackage{tikz}
\usepackage{rotating}
\hypersetup{colorlinks=true}

\begin{document}

\title{Influence of nonequilibrium phonons on the amplitude of magnetoquantum oscillations in the point-contact resistance}

\author{N.L. Bobrov$^{a,b}$, J.A. Kokkedee$^a$, N.N. Gribov$^{a,b}$, I.K. Yanson$^b$, A.G.M. Jansen$^a$}
\affiliation{$^a$High Magnetic Field Laboratory, Max-Planck-Institut f$\ddot{u}$r Festk$\ddot{o}$rperforschung and Centre National de la Recherche Scientifique,
25 avenue des Martyrs, BP 166, F-38042 Grenoble CEDEX 09, France\\
$^b$Institute for Low Temperature Physics and Engineering of the Ukrainian Academy of Sciences, 47 Lenin avenue, 310164 Kharkov, Ukraine\\
Email address: bobrov@ilt.kharkov.ua}

\published {\href{https://doi.org/10.1016/0921-4526(94)00246-R}{Physica B}, \textbf{204}, 83 (1995)}
\date{\today}

\begin{abstract}For metallic point contacts with Be and Al the magnetoquantum oscillations in the contact resistance have been investigated as a function of the applied voltage over the contact. For one set of point contacts the oscillation amplitude is found to vary nonmonotonously with the applied voltage with similarities to the point-contact spectrum of the electron-phonon interaction. The other part of the investigated point contacts shows a decrease of the oscillation amplitude with increasing bias voltage. For the understanding of the voltage dependence of the amplitude of the point-contact magnetoresistance oscillations the influence of nonequilibrium phonons generated by the ballistically injected electrons will be discussed.

\pacs {71.38.-k, 73.40.Jn, 74.25.Kc, 74.45.+c}
\end{abstract}

\maketitle
\section{Introduction}
In a magnetic field $B$ the permitted electronic states in
$k$-space are located on the so-called Landau tubes. The
cross-sections of these tubes are determined by the condition
of area quantization in a plane perpendicular to the
magnetic field direction. If one considers the tube with
the largest area of cross-section which is still partially
inside the Fermi surface (FS), its occupied part will shrink
as $B$ increases and vanishes with infinite rate at the
moment the tube touches the FS. Such an abrupt decrease
of the occupation occurs periodically in the inverse
field $1$/B causing the oscillation of free energy and magnetization
(de Haas-van Alphen effect) as $B$ changes. The
frequency $F$ of such oscillations is given by the Onsager-
Lifshitz relation

\begin{equation} \label{eq__1}
F=(\hbar /2\pi e)A,
\end{equation}
where $A$ is the extremal area of the FS cross-section.

Blurring the boundary between occupied and unoccupied
states at finite temperature $T$ reduces the amplitude
of the oscillations. The corresponding reduction
factor is
\begin{equation} \label{eq__2}
{{R}_{T}}=\frac{2{{\pi }^{2}}n{{k}_{B}}T/\hbar {{\omega }_{c}}}{\text{sh}(2{{\pi }^{2}}n{{k}_{B}}T/\hbar {{\omega }_{c}})};
\end{equation}
where $n$ is the harmonic number of the oscillations and
$\omega_c= eB/m$ the cyclotron frequency. Furthermore, due to
elastic scattering on the impurities the momentum relaxation
time of the electrons becomes finite. This leads to
the broadening of the Landau levels and accordingly
decreases the amplitude nearly in the same way as if the
temperature rises from the actual value $T$ to
$T_\text{eff} = T+x$. The corresponding reduction factor
(named Dingle factor) has the form
\begin{equation} \label{eq__3}
{{{R}_{D}}=\exp (-2{{\pi }^{2}}{{k}_{B}}nx/\hbar {{\omega }_{c}}),}
\end{equation}
where $x =\hbar/2\pi k_B\tau$ is known as the Dingle temperature.
As it follows from the theory for noninteracting particles,
the contribution to the Dingle temperature results from
any electron scattering process including electron
phonon collisions \cite{1}. However, with many-body effects
taken into account, the scattering of electrons on
phonons seems not to influence the Dingle temperature
\cite{1}. For rising temperatures, the additional electron scattering
by phonons is compensated by the decrease of
effective mass $m^*$ approaching the "bare" mass value $m$.
This is due to the fact that the reduction factor $R_{T}$ for the
temperature dependence keeps almost exactly the same
form but the "bare" mass $m$ of the electrons in Eq.(\ref{eq__2})
should be replaced by the effective mass $m^*(T)=m(1+\lambda(T))$
renormalized for the electron-phonon interaction
($\lambda$ is the electron-phonon interaction (EPI) parameter).
Although the EPI renormalizes the cyclotron mass
which determines the reduction of amplitude with temperature,
it has no influence on the mass value involved into
the Dingle factor $R_D$ related to collision broadening, i.e.
the "bare" mass $m$ has to be taken into expression (3) \cite{1}.

In traditional experiments of magnetic oscillations the
investigation of the effect of the electron-phonon interaction
on the amplitude of the oscillations is only possible
for a limited number of metals with a low Debye temperature
(e.g. Hg), where phonons may be excited at temperatures
low enough for the oscillations to be observed.
Point contacts with the size of tens or hundreds of
angstroems present a unique opportunity to study the
effect of nonequilibrium phonons generated in the contact
on the amplitude of the magnetic oscillations. The
simplest model of the contact is a circular orifice in a thin
nontransparent screen, separating two metal half-spaces.
If the contact size (diameter $d$) is small compared to the
inelastic mean free path $l_e$ of the electrons, a voltage
$V$ applied over the contact yields a nonequilibrium distribution
of the electrons with a well-defined energy $eV$. To
put it differently, the electrons are being divided into two
groups with a difference $eV$ in the Fermi level exactly
given by the applied voltage \cite{2}, In this case the electrons
with an excess energy can make a transition to a lower
energy emitting one phonon. The multi-phonon emission
processes also take place, but their probability is much
less. These emission processes lead to a voltage-dependent
increase in the contact resistance containing
detailed information on the energy dependence of the
inelastic scattering of the electrons \cite{3}. The experimental
study of nonlinear current-voltage characteristic of
point-contacts (a technique known as point-contact
spectroscopy) has been successfully applied for the direct
determination of the Eliashberg function $\alpha^2F(\omega)$ for the
electron-phonon interaction.

Thus, by the bias applied across the contact, the
nonequilibrium phonons are generated in the contact
area with any possible value of energy up to $eV$. It is of
a great importance to note that, due to $l_e> d$, only
a small fraction of the total power is released in the direct
vicinity of the contact area and the temperature of the
contacting banks of the contact remains practically equal
to that of the helium bath. With the used technique with
mechanical adjustment of the point contact the elastic
mean free path $l_i$ of the electrons in the construction area
is of the same order as the contact size $d$ which is much
less than the mean free path in the electrode. In this case
the contact resistance for zero bias in the model of a circular
orifice is determined by the Wexler formula \cite{4}
\begin{equation} \label{eq__4}
{{{R}_{0}}=\frac{16\rho {{l}}}{3\pi {{d}^{2}}}+\beta ({{l}_{i}}/d)\frac{\rho }{d}},
\end{equation}

where $\rho l= p_F/e^2n_0$ is the product of resistivity $\rho$ and
mean free path $l$ of the electrons, constant for a given
material; $p_F$ the Fermi momentum and $n_0$ the charge
carrier density, $\beta(l_i/d)$ in Eq.(\ref{eq__4}) is a monotonously decreasing
function of $l_i/d$ with $\beta(0)= 1$ and $\beta(\infty) =9\pi^2/128$. In this expression the first term describes the
Sharvin component of the point-contact resistance,
which is independent of the electron mean free path and
is determined by the topology of the FS for a given metal.
The second term is the Maxwell component of the pointcontact
(PC) resistance which depends on the metal purity
in the constriction area. When a magnetic field is
applied to a point its resistance shows an oscillating
component as a result of the quantization of the electron
energy spectrum \cite{5,6}. The oscillations of the PC resistance
have its origin in both the Maxwell and the Sharvin
part of the resistance, depending on the parameters of the
metal.

In the contacts with usual metals the Larmor radius
$r_\text{B}$ of the electrons in fields up to 20~T exceeds the typical
contact diameters (for instance, for berilium only for the
largest contacts with $d\approx$500~\AA  \ and for $B$=10~T, corresponding
to $r_\text{B}=m^*v_F/eB=$3000~\AA, the quantities become
comparable). Therefore, the magnetoquantum
oscillations in the point-contact resistance are mainly the
result of quantizing conditions in the neighbouring banks
near the constriction. The oscillations in the density of
states yield oscillations in the Sharvin component of
point-contact resistance. This constitutes an essential difference
with the observed oscillations of the magnetoresistance
in semimetal point contacts \cite{5} (where the
Maxwell component of the resistance oscillates), and
the ordinary Shubnikov-de Haas (SdH) effect in bulk
conductors, caused by quantum oscillations of the collosion
integral \cite{1}.

\begin{figure}[]
\includegraphics[width=8.5cm,angle=0]{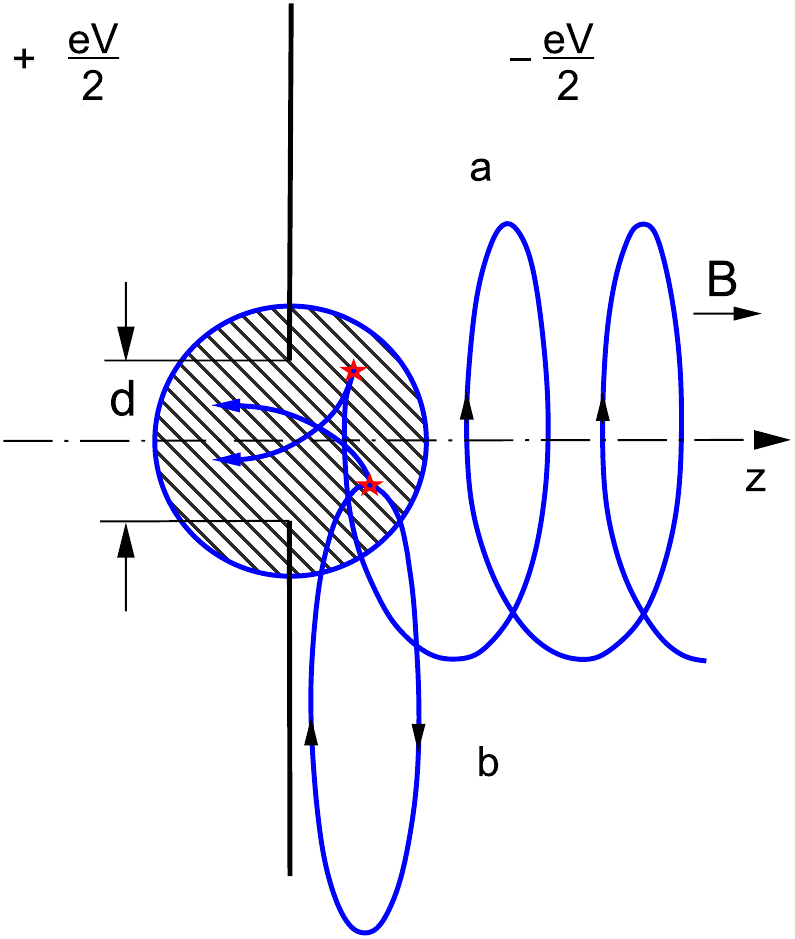}
\caption[]{Scattering of electrons moving along the quantized
orbits, by the nonequilibrium phonons (denoted by stars) accumulated
near the contact. In a scattering event an electron
changes its momentum along $z$ with on average a more effective
contribution to the current, for initial states with a small negative
$v_z$-component (a) and a zero $v_z$-component (b).}
\label{Fig1}
\end{figure}

It is quite surprising that the influence of impurities in
the constriction area is opposite to the usual dependence
as expressed by the Dingle reduction factor for bulk
material. When the magnetic field is applied parallel to
the contact axis impurities in the contact area increase
the amplitude of the magnetoresistance oscillations
\cite{7,8}. The contribution to the magnetoquantum oscillations
comes mainly from the areas of extremal crosssection
of the FS perpendicular to $p_z$ (the $z$-axis is parallel
both to the magnetic field direction and the point-contact
axis). The small value of the velocity component $v_z$ of the
charge carrier transport through the ballistic constriction
yields only quite small oscillations in the Sharvin component
of the resistance. In the presence of elastic scatterers
in the contact area the momenta of the electrons
get an isotropic distribution of directions (Fig.\ref{Fig1}), leading
to an effective increase of the $z$-component of the transport
velocity for the electrons on extremal cross-sections
and correspondingly to an increase of the magnetoquantum
oscillations in the Sharvin component of PC resistance.
The presence of elastic scatterers in neighbouring
banks causes the increase of the Dingle temperature and
thus the decrease of the oscillation amplitude.

The first experiments on the investigation of the effect
of nonequilibrium phonons on the amplitude of point-contact
magnetoresistance oscillations are described in
Refs.\cite{5,6}. In the discussion of the obtained experimental
results in Ref.\cite{5} only overheating of the electron
gas in the contact area was considered while in Ref.\cite{6}
two alternative mechanisms of damping were proposed,
namely the Dingle temperature rise due to electron-phonon scattering and the Joule heating of the contact
area.

In the present work we report on the experimental
study of the influence of nonequilibrium phonons on the
quantum oscillations in Be and Al point contacts. These
metals were chosen because of the presence of electron
pockets with small effective masses which allows to observe
the oscillations in rather weak magnetic fields. It
should be marked that in the point-contact experiments
the metallic contact is used both for the observation of
magnetoquantum oscillations in the resistance and for
the generation of nonequilibrium phonons by means of
an applied voltage over the contact.
\section{Samples and experimental details}

The studied point contacts of Al and Be were formed
between the edges of two single crystal electrodes with
the same orientation. In the case of Be the contact axis
coincided with the crystallographic $c$-axis while for Al it
was parallel to the (110)-axis. The applied magnetic field
($<10\ T$) was always parallel to the contact axis. The
electrodes were chemically polished in order to remove
the defect layer resulting from the spark cutting. Both
a technique of mechanically shifting the electrodes along
the edges \cite{9} and an electroforming method \cite{10} were
used for the contact adjustment at liquid helium temperatures.
All experiments were done at 1.3~$K$. The first and
second derivatives of the current $(I)$-voltage $(V)$ curves
were obtained by conventional modulation techniques
with data registration as a function of the applied voltage
and magnetic field. The characteristics $dV/dI(V)|_{B=\text{const}}$, $dV/dI(B)|_{V=\text{const}}$
and $d^2V/dI^2(V)$ have been measured.

\begin{figure}[]
\includegraphics[width=8.5cm,angle=0]{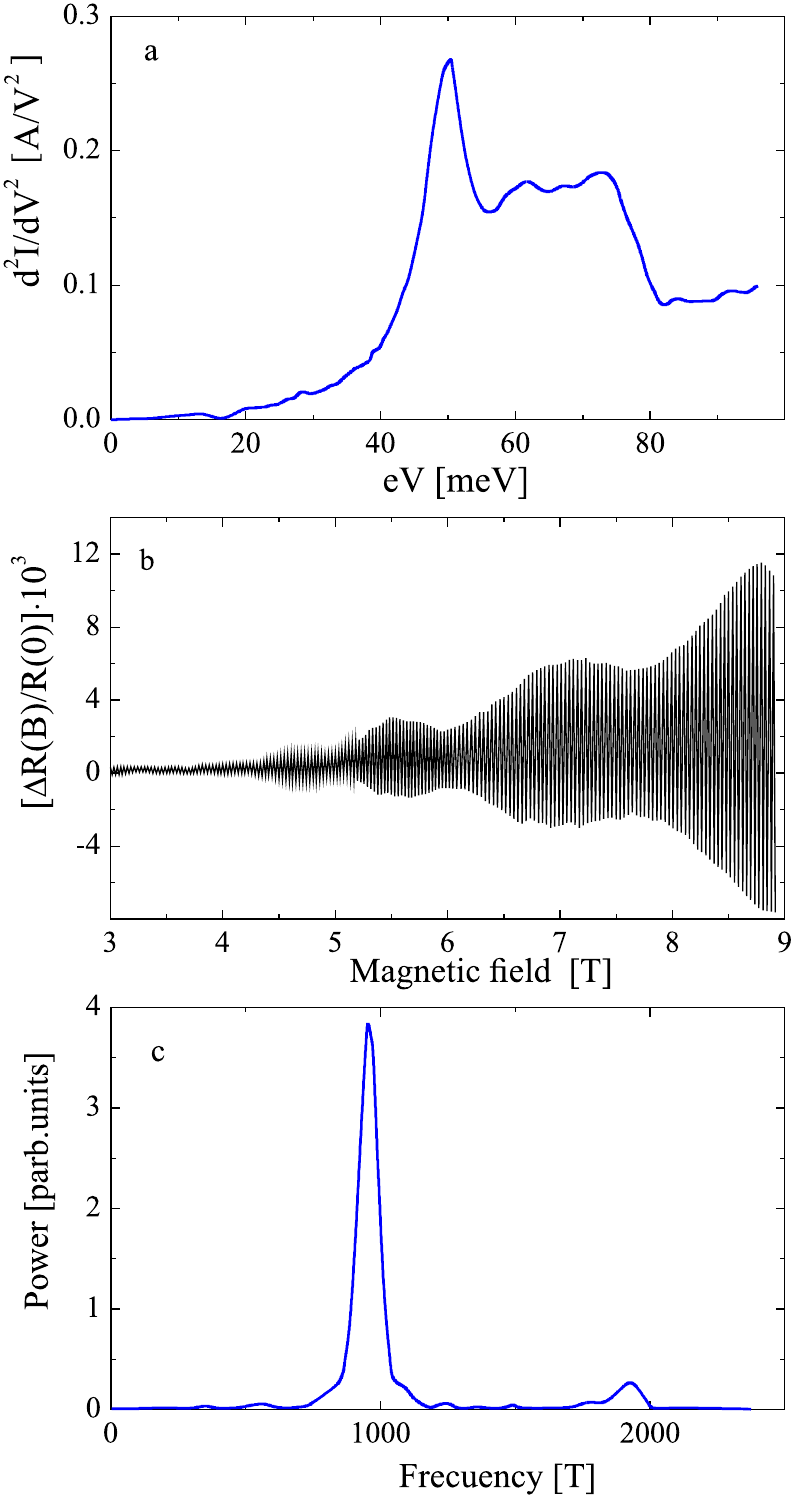}
\caption[]{(a) $d^2I/dV^2(V)$ spectrum and (b) resistance versus magnetic
field for zero bias voltage of a Be point contact of
$R=16.24\ \Omega$. The magnetic field was oriented parallel to the
$c$-axis of beryllium and parallel to the point contact axis.
(c) Fourier transform of (b) when plotted against $1/B$.}
\label{Fig2}
\end{figure}

\begin{figure}[]
\includegraphics[width=8.5cm,angle=0]{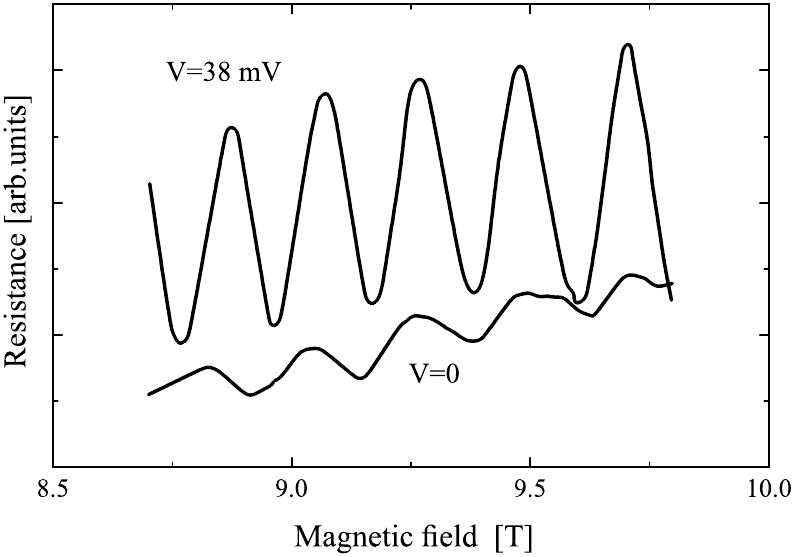}
\caption[]{Fragments of magnetoresistance oscillations for an Al
point contact of $R=141\ \Omega$ for two different bias voltages.}
\label{Fig3}
\end{figure}

\begin{figure}[]
\includegraphics[width=8.5cm,angle=0]{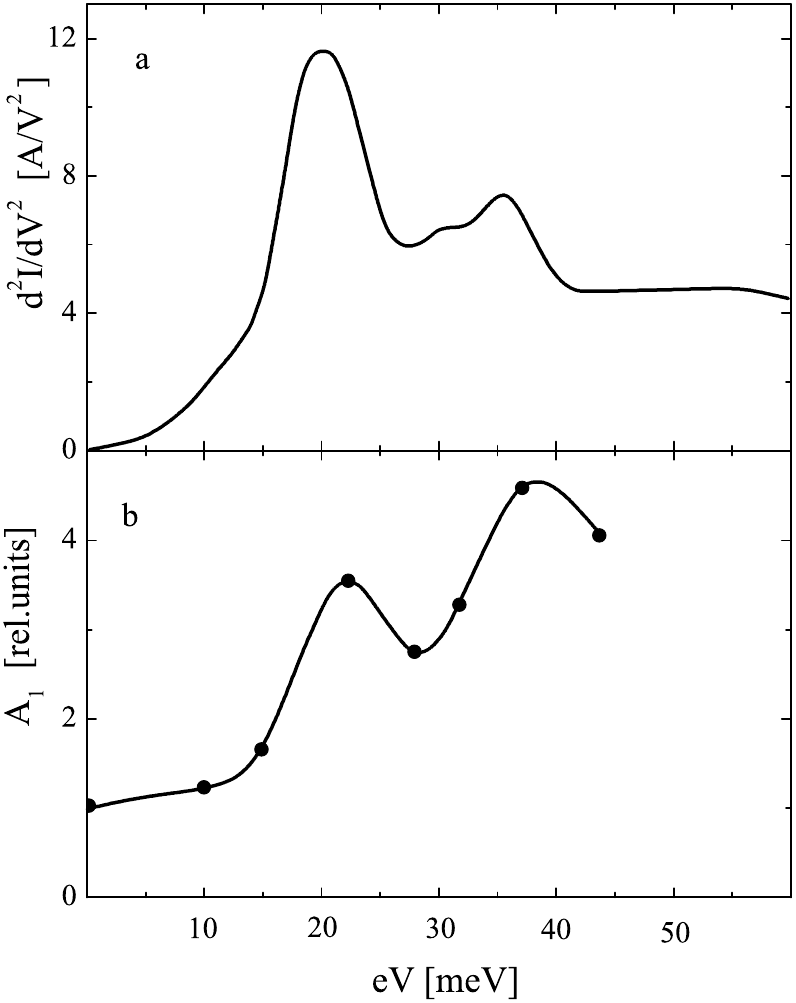}
\caption[]{(a) $d^2I/dV^2(V)$ spectrum of an Al point contact of
$R=1,41\ \Omega$. (b) Relative magnitude of the first harmonic of the
point-contact magnetoresistance oscillations versus bias voltage.
At $V=0\ mV$ and $B=9.5\ T$, $\Delta R/R_0\approx 5\times 10^{-4}$.}
\label{Fig4}
\end{figure}

\begin{figure}[]
\includegraphics[width=8.5cm,angle=0]{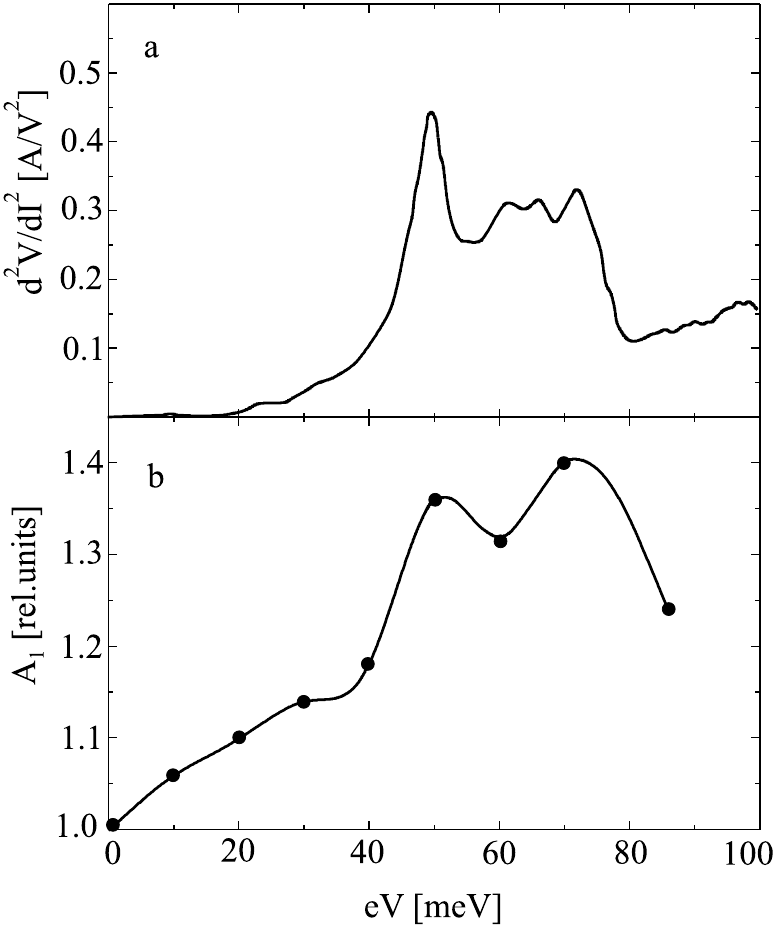}
\caption[]{(a)$d^2I/dV^2(V)$ spectrum of a Be point contact
of $R=8.66\ \Omega$. (b) Relative magnitude of the first harmonic
of the point-contact magnetoresistance oscillations versus
bias voltage. At $V=0\ mV$ and $B=9.5\ T$, $\Delta R/R_0\approx 2\times 10^{-3}$.}
\label{Fig5}
\end{figure}

\begin{figure}[]
\includegraphics[width=8.5cm,angle=0]{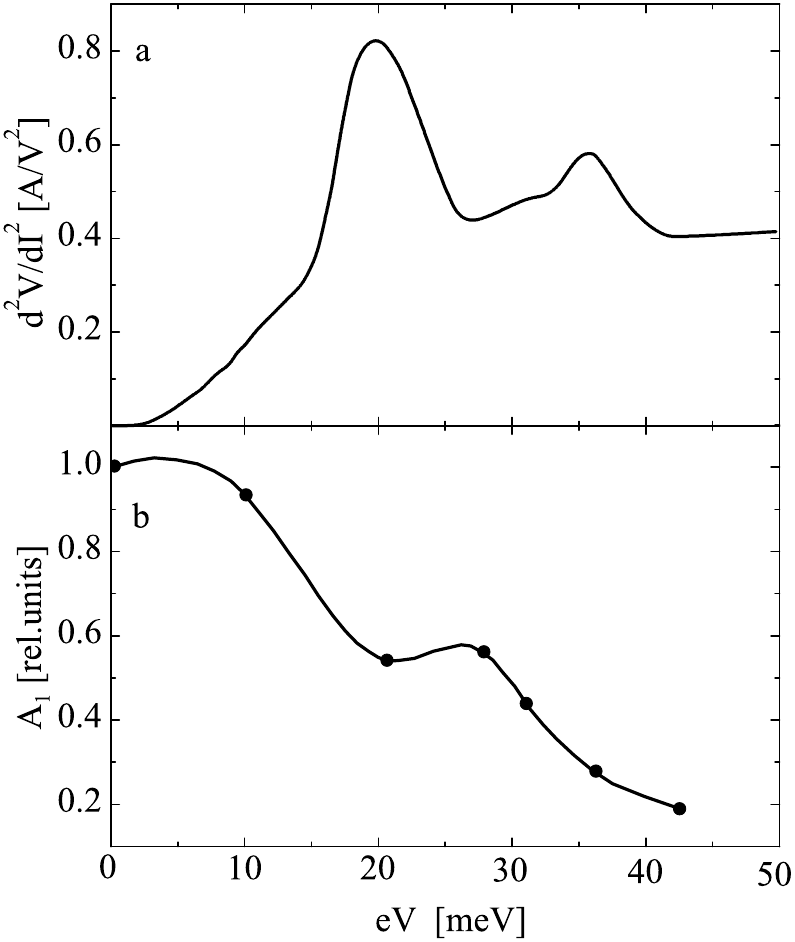}
\caption[]{(a) $d^2I/dV^2(V)$ spectrum of an Al point contact of
$R=8.73\ \Omega$ (b) Relative magnitude of the first harmonic of the
point-contact magnetoresistance oscillations versus bias voltage.
At $V=0\ mV$ and $B=9.5\ T$, $\Delta R/R_0\approx 3\times 10^{-3}$.}
\label{Fig6}
\end{figure}
Fig.\ref{Fig2} presents an example of the $d^2I/dV^2\equiv-(d^2V/dI^2)/(dV/dI)^3$ spectrum of a Be contact of $16.24\ \Omega$, the oscillating
component of magnetoresistance and its Fourier
transformed spectrum. The spectrum shows clearly the
structure between 40 and 80~$meV$, which can be related to
the $\alpha^2F(\omega)$ function for the electron-phonon interaction.
The magnetoquantum oscillations studied in the Be resistance
correspond to third-band electron surface (cigar) with
frequency 985~$T$ as known from de Haas-van Alphen
experiments \cite{11}. For various bias voltages across the
contact the oscillations of $R(B)=dV/dI(B)$ were recorded
in the same magnetic field interval (as a rule in the
range between 8.5 and 10~$T$). Fig.\ref{Fig3} presents the fragments
of such records for two different bias voltages of an Al-Al
point contact. For the Al contacts the investigated oscillation
corresponds to the third zone $\gamma$-pocket with frequency
290~$T$ \cite{12}. After the recording of the field dependent
curves for several bias voltages, the record for $V = 0$ was
compared with the initial one in order to check the contact
stability during the measurements.

Contacts with different resistances (from 0.6 to 20~$\Omega$,
background levels (from 30\% to 85\%) in the high voltage
part of the $d^2I/dV^2(V)$  spectrum and degrees of point-contact
spectrum broadening were studied. For metals
like Be and Al the given resistance values correspond to
contact diameters ranging from 50 to 500~\AA. A full set of
characteristics for the voltage and magnetic field dependences
could be obtained for five Al and five Be contacts.
In the given magnetic field interval the amplitude of the
oscillations was obtained by means of a Fourier analysis.
Since the $dV/dI(B)\mid_{V=\text{const.}}$ curves were measured at
a constant modulation current while the differential resistance
increases with applied bias (increase up to 10\% in
the Debye energy range), the obtained amplitudes were
normalized for a fixed modulation voltage (the value
measured at zero bias). Figs. \ref{Fig4}-\ref{Fig6} present typical results
for the measured contacts. For one group of the investigated
point contacts the oscillation amplitude $A_1$ of the
PC magnetoresistance increases with increasing bias and
its voltage dependence is similar to the point-contact
spectrum of the electron-phonon interaction (Figs.\ref{Fig4} and
\ref{Fig5}). Some of the aluminium contacts exhibit a nonmonotonic
decrease of the amplitude as a function of voltage (Fig.\ref{Fig6}). In the following we discuss the possible reasons
of such dependencies.
\section{Discussion}

The relaxation of the injected electrons in the point-contacts
leads to the generation of phonons via spontaneous
emission processes. The resulting nonequilibrium
distribution of the phonons will depend on the applied
voltage over the contact. In analogy with the discussed
influence of elastic impurity scattering, the scattering of
the electrons with these nonequilibrium phonons will
influence the amplitude of the magnetoresistance oscillations.
As for the elastic scattering, this phonon-induced
scattering has a quite different effect on the quantum
oscillations for scattering processes in the centre of the
contact (with resulting increase of amplitude) compared
with scattering in the banks of the constriction (with
resulting decrease of amplitude). In the following we will
only give a qualitative discussion of the generated
phonon distribution and its influence on the magnetoquantum
oscillations.
Although the elastic mean free
path $l_i$ of the electrons is of the order of the contact
diameter $d$ for mechanically adjusted point contacts, the
diffusion length $\Lambda_e =(l_il_e)^{1/2}$ for inelastic relaxation is
mostly larger than the contact dimension. At the Debye
energy the ballistic mean free path $l_e$ for inelastic scattering
equals approximately 3000~\AA  \ for aluminium and
1500~\AA  \ for beryllium.
\begin{figure}[]
\includegraphics[width=8.5cm,angle=0]{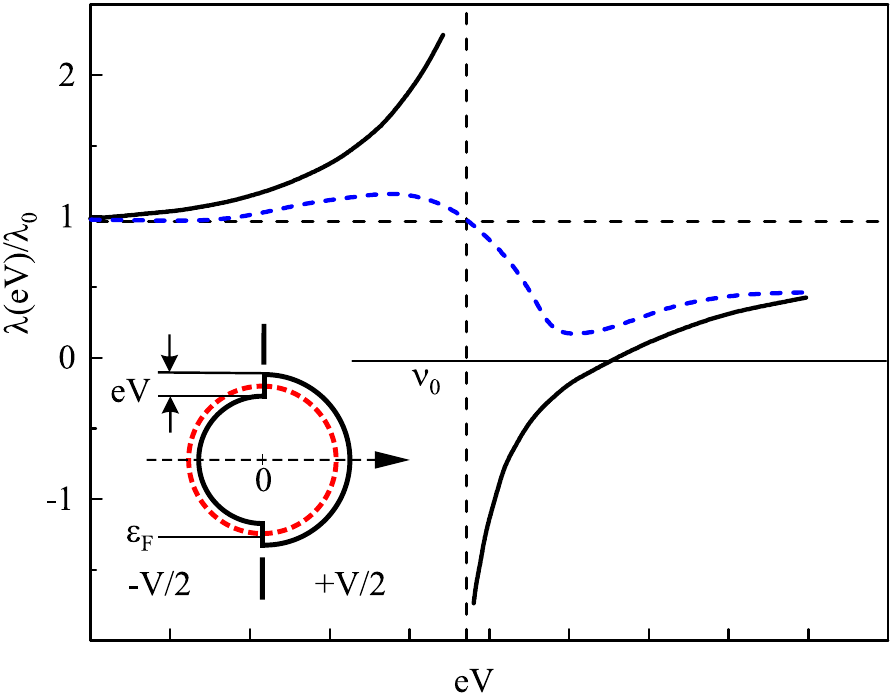}
\caption[]{Schematic representation of theoretically predicted \cite{14}
behaviour of the electron-phonon interaction parameter $\lambda$ versus
voltage bias near the orifice. Solid and broken lines correspond,
respectively, to the infinitesimal small and finite widths of
the phonon band at the characteristic energy $eV =\nu_0$. The inset
shows an electron distribution function of a ballistic point contact
in the centre of the orifice.}
\label{Fig7}
\end{figure}

Apart from the impurities that have a weak influence
on the elastic mean free path $l_i^{\text{ph}}$ of the phonons although
being effective scatterers for the electrons, the pressuretype
contacts contain as a rule a great number of lattice
defects. These defects have little influence on the elastic
mean free path of electrons $l_i$, but reduce significantly the
elastic mean free path $l_i^{\text{ph}}$ of the nonequilibrium phonons
generated by the flow of electrons through the constriction
\cite{13}. If $l_i^{\text{ph}} < d$, the additional electron scattering by
the nonequilibrium phonons causes a background in the
point-contact spectrum (voltage independent value of the
second derivative curves $d^2I/dV^2(eV)$ at $eV > \hbar\omega_D$)
comparable with the spectral part of the point-contact
spectrum of EPI. In this case a nonequilibrium distribution
of phonons is established in the contact with an
effective temperature dependent on the position with
respect to the contact and on the reabsorbtion coefficient
of phonons in the contact area.

The nonequilibrium distribution of the phonons in
a point contact differs from Planck's one by the existence
of a cut-off at $\hbar\omega_D=eV$ (for details see Ref.\cite{13}). In this
description of the nonequilibrium distribution one can
define a diffusive mean free path $\Lambda_e^{\text{ph}}=(l_i^{\text{ph}}l_e^{\text{ph}})^{1/2}(\omega_D/\omega)^{1/2}$ for the distance over which the nonequilibrium
phonon looses its energy. Here $l_e^{\text{ph}}$ is the ballistic inelastic
mean free path of the phonons. For Debye energies
$l_e^{\text{ph}}$ equals by order of magnitude the inelastic mean free
path $l_e$ of the electrons and has the same energy dependence,
i.e. decrease of $l_e^{\text{ph}}$ at the characteristic phonon
frequencies. The high-frequency phonons having the
smallest $\Lambda_e^{\text{ph}}/d$ ratio are accumulated near the orifice and
strongly interacting with the electrons. The low-frequency
phonons (larger $\Lambda_e^{\text{ph}}/d$ ratio) dominate in the
banks. Assuming the exact analogy with the impurity-scattering,
the scattering of the electrons in the contact
area causes the randomization of the momenta of the
electrons. In this case the current distribution of the
$z$-component of the velocity at extremal cross-sections of
the Fermi surface increases in a magnetic field leading to
an increase of the magnetoquantum oscillations. The
electron scattering with generated phonons arriving in
the banks leads to a reduction of the amplitude of the
oscillations.
The resulting voltage dependence of the amplitude
of the point-contact magnetoresistance oscillations
is determined by the balance between these two
processes.

For high Ohmic contacts with aluminium ($R>5\ \Omega$,
$d<150$~\AA) the decrease of the oscillation amplitude with
increasing bias voltage is always observed. These contacts
are small compared to the phonon mean free path
and all phonons are leaving for the banks before scattering
with electrons in the contact centre takes place. The
electron-phonon scattering in the banks reduces the
amplitude of the oscillations. For low Ohmic contacts
with aluminium the inequality $\Lambda_e^{\text{ph}}>d$ is not so rigorous
and the phonons accumulate in the contact area. The
resulting scattering with electrons in the contact region
leads to the observed increase of the oscillation amplitude.
One would expect a correlation between the increase
of the magnetoresistance oscillations and the occurrence
of a pronounced background signal in the second
derivative $d^2V/dI^2(V)$ spectra due to the accumulation
of nonequilibrium phonons in the contact area.
Unfortunately, we have not seen such a correlation in the
data.

As for beryllium, the increase of the amplitude of
magnetoresistance oscillations with bias rise was observed
for all contacts. It is probably connected with
extremely high rigidity of the Be crystal lattice. As a consequence,
the deformations formed during contact creation
would be stopped near the surface and do not
spread to the bulk material thus creating efficient reflectors
concentrating the phonons in the contact.

The given qualitative model for the voltage dependence
of the point-contact oscillations explains only the
monotonous part of the voltage dependence but not
its similarity to the structure in the EPI spectrum. The
phonon energy relaxation length $\Lambda_\omega$ will decrease
stepwise at voltages corresponding to the phonon frequencies
(alike the electron energy relaxation length $\Lambda_e$).
Therefore, the voltage dependence of the amplitude
should be expected to be similar to the continuous first
derivative curve $dV/dI(V)$. In the following we discuss
the energy dependence of the mass-renormalization for
the nonequilibrium electrons in a point contact as a possible
reason for the nonmonotonous structure observed.

It was predicted in Ref.\cite{14} that the electron mass
renormalization parameter $\lambda$ for the effective mass
$m^* = m(1 + \lambda)$ differs from the equilibrium value $\lambda_0$ in
the vicinity of the orifice. For an applied voltage over
a ballistic contact the resulting nonequilibrium distribution
of the electrons differs from the spherical Fermi
sphere that would give the equilibrium value $\lambda_0$ for the
electron-phonon enhancement factor. For the energy
distribution of a ballistic contact (see inset of Fig.\ref{Fig7} for
this distribution in the centre of the contact) $\lambda$ will depend
on the applied voltage. In the centre of the contact
one finds \cite{14}
\begin{equation} \label{eq__5}
\lambda (eV)=\frac{{{\lambda }_{0}}}{2}\left[ {1+\nu _{0}^{2}}/{\left( \nu _{0}^{2}-{{\left( eV \right)}^{2}} \right)}\; \right]
\end{equation}

for an Einstein model of the phonon distribution with
phonon energy $\nu_0$. The expected behaviour of $\lambda(eV)$ is
shown schematically in Fig.\ref{Fig7}. It can be seen that for high
voltages the mass renormalization has been reduced by
a factor 2 because only half of the Fermi sphere takes part
in the mass renormalization for the nonequilibrium distribution
in a point contact (see inset of Fig.\ref{Fig7}). For
$eV=\nu_0$ a resonance phenomenon occurs in the mass
renormalization. In the same figure we have also given
the result for $\lambda(eV)$ for a broadened band of phonon
energies around $\nu_0$. From the theory can be concluded
that the nonmonotonous energy dependence of the
effective electron mass at the characteristic phonon frequencies
is inherent to the nonequilibrium electron
distribution function in point contacts under finite bias
condition.

The energy-dependent mass enhancement in a point
contact could cause the observed nonmonotonous dependence
of the magnetoquantum oscillation amplitude
on voltage bias. Note from Fig.\ref{Fig7} that the inflection point
with negative slope on the $A_1(eV)$-curve should be taken
as the corresponding characteristic phonon energy. Although
the exact position of the maxima in the voltage
dependence of the oscillation amplitude changes slightly
form contact to contact, the voltage position of these
inflection points are usually located at biases larger than
the maxima in the phonon density of states. The theory of
the voltage-dependent mass enhancement has been
evaluated in the absence of a magnetic field \cite{14}. It
would be very interesting to include the energy dependence
of the effective mass in the theories explaining the
observation of magnetoquantum oscillations in ballistic
point contacts. We have probably observed such an
energy dependence of the effective mass enhancement in
the voltage dependence of the oscillation amplitude in
a ballistic PC. The ballistic electron transport in a point
contact has the advantage to study the energy dependence
of the effective mass in the absence of thermal
phonon smearing.
 \section{Conclusion}
 In the present work we have studied the voltage dependence
of the magnetoquantum oscillation amplitude
in the resistance of a metallic point contact in a magnetic
field parallel to the contact axis. For the first time it was
found that such a dependence shows a nonmonotonous
increase with structures similar to the point contact spectrum
for the electron-phonon interaction. Such an effect
was observed for berilium point contacts as well as for
low ohmic aluminium contacts. The possible reason for
the increase of the oscillation amplitude may be due to
the additional scattering of Landau quantized electrons
by phonons generated by the accelerated electrons. These
scattering processes of electrons with nonequilibrium
phonons enhance the contribution of the charge carrier
transport through the contact for the electrons at extremal
Landau orbits. Therefore, opposite to the usual
suppression of magnetoquantum oscillations by impurity
scattering in a bulk metal, the oscillations increase with
the increasing scattering at applied voltages. If the scattering
with nonequilibrium phonons extends to the
banks of the contact the well-known decrease of oscillation
amplitude is observed (for high ohmic aluminium
contact). The nonmonotonous $eV$-dependence (i.e. presence
of peaks near the maxima in the phonon density of
states) is possibly connected with the energy-dependent
renormalization of the electron mass in the nonequilibrium
distribution of the electrons in a ballistic contact as
defined by the applied voltage.

\section{Acknowledgements}
The authors would like to thank A. N. Omel'yanchuk,
E. N. Bogachek, A. A. Zvyagin and A. D. Dan'kovsky for
stimulating discussions during the preparation of this
paper.

\end{document}